# Completely Spin-Decoupled Geometric Phase of Metasurface

Xinmin Fu[1,2,6], Jie Yang[3,6], Jiafu Wang[1,2,4], Yajuan Han[1,2,5], Chang Ding[1,2], Tianshuo Qiu[1,2], Bingyue Qu[3], Lei Li[1,2], Yongfeng Li[1,2], Shaobo Qu[1,2]

1. Department of Basic Sciences, Air Force Engineering University, Xi'an 710051, China
2. Shaanxi Key Laboratory of Artificially-Structured Functional Materials and Devices, Xi'an 710051, China
3. School of Electronic and Information Engineering, Xi'an Jiaotong University, Xi'an, 710049, China
4. e-mail: wangjiafu1981@126.com
5. e-mail: mshyj_mail@126.com
6. These authors contributed equally to this work.

**Abstract**：Metasurfaces have provided unprecedented degree of freedom (DOF) in manipulating electromagnetic (EM) waves. Geometric phase can be readily obtained by rotating the meta-atom of metasurfaces. Nevertheless, such geometric phases are usually spin-coupled, with the same magnitude but opposite signs for left- and right-handed circularly polarized (LCP&RCP) waves. To achieve independent control on LCP and RCP waves, it is crucial to obtain spin-decoupled geometric phases. In this paper, we propose to obtain completely spin-decoupled geometric phases by engineering surface current paths on meta-atoms. Based on the rotational Doppler effect, the rotation manner is firstly analyzed and it is found that the essence of generating geometric phase lies in the rotation of surface current paths on meta-atoms. Since the induced surface currents paths under LCP and RCP waves always start oppositely and are mirror-symmetrical with each other, it is natural that the geometric phases be with the same magnitude and opposite signs when the meta-atoms are rotated. To obtain spin-decoupled geometric phases, the start point of induced surface current under one spin should be rotated by an angle while that under the other spin by another different angle. In this way, LCP and RCP waves can acquire different geometric phase changes and spin-decoupled geometric phase can be imparted by metasurfaces. Proof-of-principle prototypes were designed, fabricated and measured. Both the simulation and experiment results verify spin-decoupled geometric phases. This work provides a robust means of obtaining spin-dependent geometric phase and will further adds up to the metasurfaces'DOF in manipulating EM waves.

## 1. Introduction

Freely customizing the properties and functions of available materials is the persistent pursuit of scientists, scholars and engineers. Metasurfaces, planar arrays of subwavelength metallic or dielectric structures (meta-atoms), have provided a robust recipe for customizing materials since they allow unprecedented degree of freedom (DOF) in manipulating electromagnetic (EM) waves [1-4]. By changing the geometrical parameters or orientations of meta-atoms, the amplitude [5], phase [6,7] and polarization [8] of the EM waves can be tailored, so as to implement functional devices such as planar focusing lenses [9], planar retroreflectors [10,11] and holographic imagers [12,13]. Among them, metasurfaces have played very important roles in phase control and different kinds of phases can be obtained using metasurfaces.

In general, the phases imparted by metasurfaces mainly includes four types:

resonance phase, propagation phase, detour phase and geometric phase. The resonance phase is resulted from the localized electron oscillation within meta-atoms and are typically Lorentzian type [14]. The propagation phase is resulted from phase accumulation within the very thin dielectric substrates of meta-atoms [6,15]. The detour phase only appears under oblique incidence and is strongly dispersive both in frequency domain and in spatial domain[16,17]. The above three kinds of phases are all in nature resulted from localized resonances within or between meta-atoms. Therefore, they are all in nature narrowband, which restricts their applications in developing broadband or even wideband functional devices. The geometric phase is usually obtained by rotating meta-atoms by sequences. Such phases are also known as Pancharatnam-Berry (P-B) phases and are usually nondispersive in a quite wide band[18-21]. Geometric phases have brought great impetus in developing wideband functional devices using metasurfaces, due to the wideband properties. Nevertheless, there is still an underlying dilemma for geometric phases obtained by rotating meta-atoms. According to the generating mechanism of geometric phases, the rotated meta-atom always imparts geometric phases with the same amplitude and opposite signs for left- and right-handed circularly polarized (LCP&RCP) waves and such geometric phases are usually spin-coupled [6,21], which prohibits independent controls on LCP and RCP waves. In order to overcome this limitation, a common method is to combing the spin-coupled geometric phase with the spin-decoupled phase. In this way, the magnitudes or even the signs of geometric phases imparted to LCP and RCP can be changed. For example, bifunctional wavefront-manipulating devices are achieved by combining geometric phase and propagation phases [6,15]. Nevertheless, this method depends on introduction of a 'third-party' phase, and do not break the spin-symmetry of geometric phases in nature. The geometric phases imparted to LCP and RCP waves are in fact still spin-coupled. Many recent works have proposed other spin-decoupling methods, such as resonance phase and P-B phase[22-25], Aharonov-Anandan (A-A) phase and P-B phase [26,27], also by combining geometric phase with a 'third-party' phases.

Then, it comes naturally that can we obtain spin-decoupled geometric phases using metasurface without combinations? In this paper, we explore spin-decoupled geometric phase of metasurface and propose to obtain spin-decoupled geometric phases by engineering the surface current paths on meta-atoms. To this end, the geometric phase obtained by rotating meta-atoms is firstly analyzed. It is found that the geometric phase of rotation is due to the angular Doppler frequency shift and that the essence lies in the rotation of surface current path on meta-atoms. Since the induced surface currents paths under LCP and RCP waves always start oppositely and are mirror-symmetrical with each other, it is natural that the geometric phases be with the same magnitude and opposite signs when the meta-atoms are rotated, as is shown in Fig 1(a). Within this in mind, to obtain spin-decoupled geometric phases, it is expected that the start point of induced surface current under one spin should be rotated by an angle while that under the other spin by another different angle. In this way, LCP and RCP waves can acquire different geometric phase changes from metasurface and spin-decoupled geometric phase can be achieved using such metasurface. As a proof of concept, we use the split-ring resonator (SRR) structure as the meta-atom to demonstrate this method. By changing the starting point of the opening gap, rotation of surface current on SRR structures can be enforced, without rotating the SRR meta-atom. Due to the actual rotation of surface current path, geometric phase is naturally generated. As is depicted in Fig 1(b), by separately customizing the start points of surface currents induced by LCP and RCP waves, completely spin-decoupled geometric phases can be imparted to LCP and RCP waves by the metasurface. Prototypes were designed, fabricated and measured. Both the simulation and experiment results verify spin-decoupled geometric phases and the spin-decoupled geometric phase can be readily combined with other phases such as rotation geometric phase. This work provides a robust method of obtaining spin-decoupled geometric phases and will further adds up to the metasurface′s DOF in manipulating EM waves.

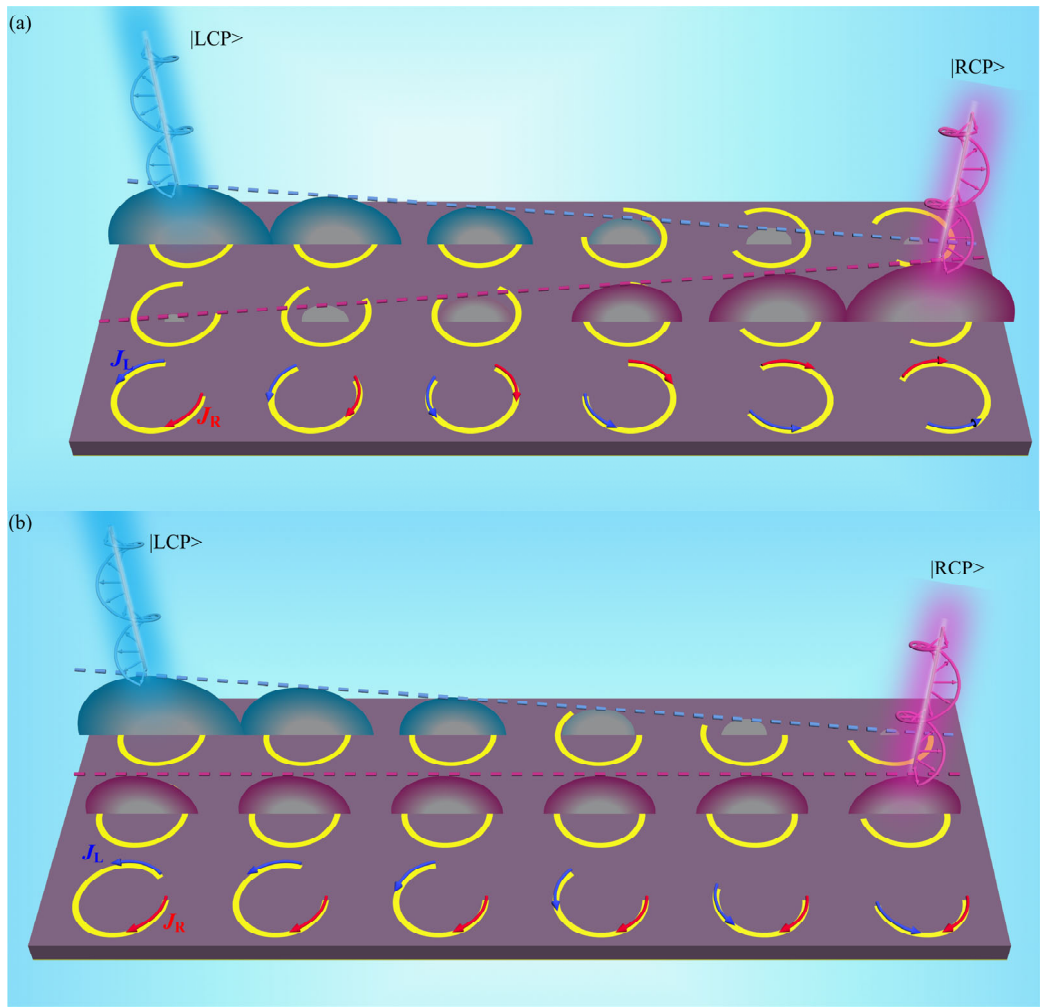

Fig 1 The two geometric phase modulation and their surface currents. The blue arrow represents the surface current $J_L$ induced by the LCP, while the red arrow represents the surface current $J_R$ induced by the RCP. (a) the spin-coupled P-B phase. (b) the spin-decoupled geometric phase.

## 2. Theory Analysis and Design

Geometric phase plays an important role in developing wideband metasurfaces due to its wavelength independent and non-dispersive properties. Under circularly polarized incidence, near-unity reflection amplitude can be maintained while wideband nondispersive reflection phase can be obtained via metasurfaces. The conventional way of obtaining geometric phase is by rotating the meta-atoms, that is, $\varphi_{L(R)}=\pm 2\alpha$, where $\alpha$ is the rotating angle, ± represents LCP and RCP waves, respectively.

We choose the simplest dipole structure (short straight metallic strip, as shown in Fig 3) to analyze the essential reason of generating geometric phases by rotation. Previous studies have shown that the rotational Doppler shift occurs when a light beam carrying angular momentum propagates through a spinning object along its rotation axis and the Coriolis effect is responsible for it. Moreover, the angular frequency resulted from the rotational Doppler shift will produce a geometric phase [28-30] (More details of the derivation see S1, Supporting Material).

$$\phi = \int \Delta\omega dt = \int \sigma \Omega_z dt \qquad (1)$$

where $\Delta\omega$ is the angular frequency shift resulting from Angular Doppler Effect, $\Omega_z$ is the rotational angular velocity with respect to the dipole, $\sigma=\pm 1$ represent LCP(-) and RCP(+) waves. The rotation of the dipole around the center represents the rotation of the polarization state, corresponding to the rotation of the polarization ellipse in the Poincare sphere depicted in Fig 2. Each longitude represents an evolutionary state of CP conversion(Fig 2(b) depicts eight of these states)[31]. Thus, the angular velocity of the rotation of the coordinate system is transformed into the rotation velocity of the polarization ellipse, i.e. $\Omega_z = \frac{d2\tau\chi}{dt}$, where $\tau=\pm 1$ represent the rotation direction of the polarization ellipse(clockwise rotation notated as 1, counterclockwise rotation notated as -1). Then Eq (1) becomes:

$$\phi = 2\sigma\tau\chi = 2\delta\chi \qquad (2)$$

where $\delta=\pm 1$. When the rotation direction is consistent with the spin direction of the incident waves, $\delta=1$; when they are opposite in directions, $\delta=-1$.

Under the illumination of CP waves, surface current will be induced on the meta-atom, which flows along the dipole, as is shown in Fig 3. The secondary radiation of the surface current actually determines the EM properties of reflected fields. Thus, the P-B phase generate by rotation is essentially caused by the rotation of the surface current. Intuitively, the surface current induced by incident waves with different spin directions (LCP or RCP) will start and flow in opposite directions, as shown in Fig 3(a). In this case, when the metallic strip is rotates by an angle $\alpha$, the surface induced by LCP waves will also be rotated by an angle $\alpha$ relative to the spin of incident waves, while the surface current induced by RCP waves will be otherwise rotated by an angle $-\alpha$ relative to the spin of incident waves. This feature is also known as spin-coupling of P-B phases. This can also be verified by the simulation results in Fig 3(a). Moreover, we find that when the meta-atom is rotated, the surface current at each point on the structure is also rotated around the center, as illustrated in Fig 3(a). Therefore, we can convert the straight strip structure into an SRR of radius $r$, as shown in Fig 3(b). For the SRR structure, the induced surface currents under LCP and RCP waves always start from the two ends of the opening gap of SRR, respectively; and they start oppositely and they also flow oppositely. When the SRR meta-atom is rotated, the induced surface currents are also rotated accordingly. The reflection phase is also given in Fig 3(b) and is found to be consistent with that of the short straight strip structure.

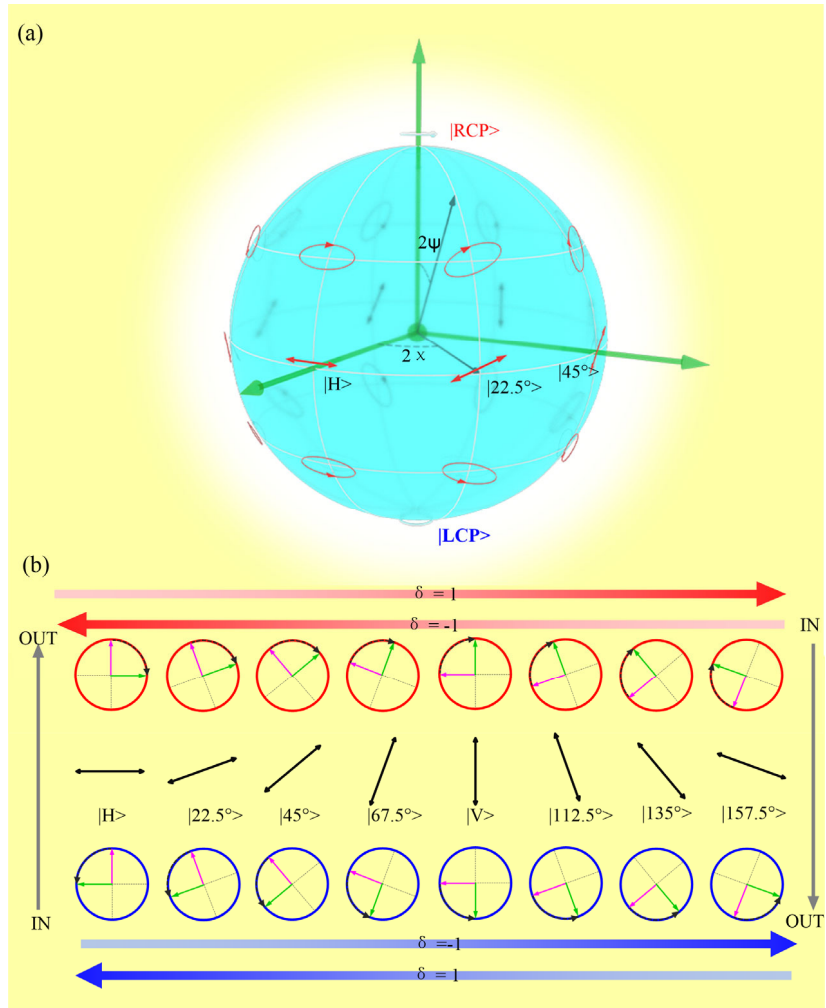

Fig.2 Geometric phase by rotation: (a) The schematic diagram of the Poincare sphere. (b) The Schematic diagram the CP conversion evolutionary state and geometric phase gradient direction.

All the simulated and analysis indicate that the geometric phase is actually caused by the rotation of surface current on the meta-atom. The core of metasurface is the metallic structure pattern design, which determines the detailed surface current path. This is why metasurface can be customized for EM manipulation. This tells us that if the rotation of surface current can be achieved by structural design without rotating the meta-atom, geometric phase can also be obtained. The current diagram in Fig 3(b) shows that the surface current flows at the two endpoints of the SRR gap as the starting point, and we can change the starting point of one path to force rotation of surface current induced under a given CP waves. For example, we can change the start point of surface current induced by one spin state while the start point of surface current induced by the other spin state keeps unchanged. In this way, geometric phase gradient can be formed for the former spin state while not for the other spin state. With this idea in mind, we firstly design a metasurface which carries geometric phase gradient only for LCP waves, but not for RCP waves. In this regard, the SRR shown in the Fig 4(a) is adopted as the meta-atom for spin-decoupled geometric phase modulation. Metallic SRR with an opening gap is patterned on a grounded dielectric substrate. F4B microwave laminate ($\varepsilon_r$=2.65, tan$\delta$=0.001) is selected as the dielectric substrate and other geometric parameters are also given in the Fig 4(a). The reflection performance of the meta-atom was firstly simulated. Circularly-polarized (CP) waves are incident along the *z* direction, while *x* and *y* directions are set as periodic boundary conditions. As shown in Fig 4(b), high-efficiency polarization conversion can be achieved within 8.0-14.0GHz. Such polarization conversion is in fact resulted from the reversed propagation direction upon reflection. The near-unity reflection magnitude of the SRR meta-atom guarantees the design of high-efficiency phase-modulation metasurface.

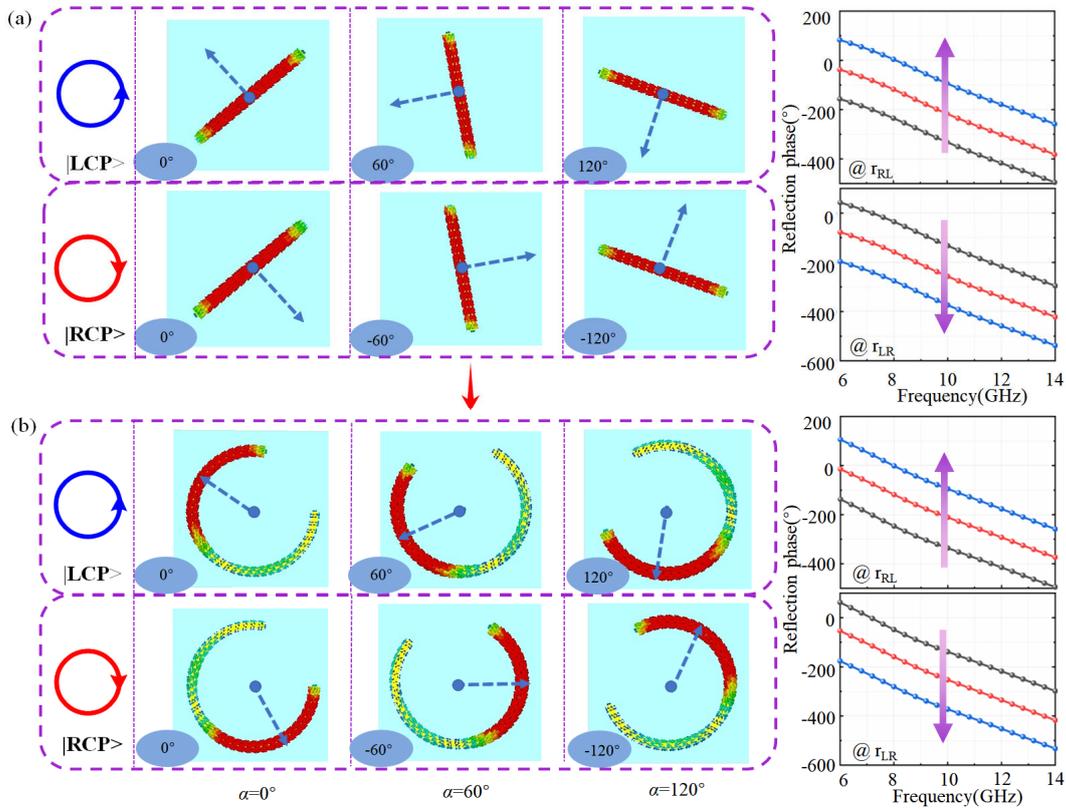

Fig.3 Geometric phases generated by rotating meta-atom(α is defined as the counterclockwise rotation angle of the meta-atom): (a) rotation of surface current on the short straight strip and the corresponding reflection phases; (b) rotation of surface current on SRR and the corresponding reflection phases

According to the analysis in Fig 3, when CP waves are incident onto the SRR meta-atom, surface currents will be induced, which start at the two ends of the opening gap, respectively, for LCP and RCP waves. As shown in Fig 4(c), surface current diagrams under different $\beta_1$(40°, 85°, 130°, 175°) are given, where $\beta_1$ and $\beta_2$ are the angle between the upper and lower ends of the opening gap referenced to the *x*-axis, respective, and they determine the start points of surface currents induced by incident LCP and RCP waves, respectively. It can be found that as $\beta_1$ changes, the surface current induced by LCP waves is forced to undergo rotation with respect to the circle center of SRR. The rotation angle of surface current patch can be expressed as $\chi=-(\beta_1-40)$, which will result in a reflection geometric phase $\varphi=2\delta\chi=2(\beta_1-40)$ (the corresponding reflection amplitude is depicted in Fig S1, Supporting Material). In contrast, the surface current induced by RCP waves does not rotate because $\beta_2$ remained constant. The reflection phase versus $\beta_1$ is also given in Fig.4(c). It can be found that as $\beta_1$ changes, reflection phase for LCP waves can cover the span of $2\pi$, which is completely consistent with the theoretical predictions, while the reflection phase RCP waves remains unchanged. When $\beta_1$ is kept unchanged and $\beta_2$ changes, RCP waves will obtain geometric phase gradient while LCP waves will not (See more details in Fig S2, Supporting Material). Therefore, by engineering the surface current path on the meta-atom, spin-independent geometric phase can be obtained, which will facilitate independent controls on LCP and RCP waves.

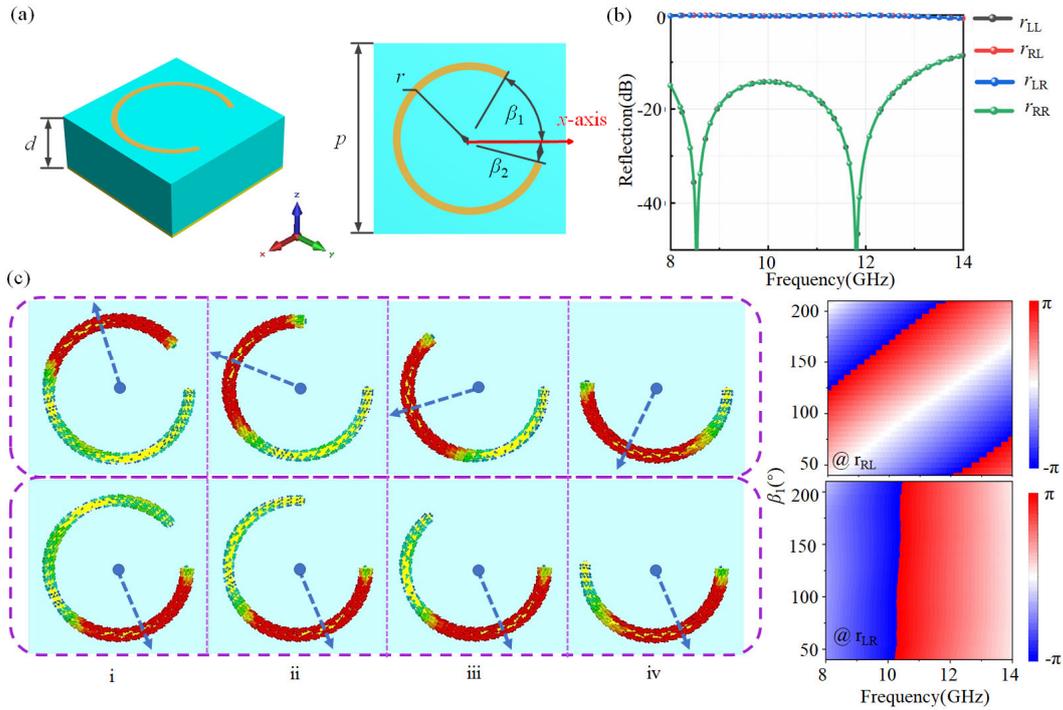

Fig.4 Spin-decoupled geometric phase obtained by engineering the surface current path on meta-atoms: (a) schematic diagram of the SRR meta-atom, where $P$=10mm, $d$=4mm, $r$=4mm, $\beta_1$=80°, $\beta_2$=0°. (b) reflection performances under incident LCP and RCP waves. (c) surface currents and reflection phases under different $\beta_1$: (i) $\beta_1$=40°, (ii) $\beta_1$=85°, (iii) $\beta_1$=130°, (iv) $\beta_1$=175°.

## 3. Spin-Decoupled Metasurface and Experimental Verification

In the last section, we have shown, by the example of SRR meta-atom, that by engineering the surface current path, different geometric phases can be imparted to LCP and RCP waves. When only $\beta_1$ changes, only LCP waves can acquire geometric phase changes ($\beta_1$-geometric phase, for short) while RCP waves does not. Similarly, when only $\beta_2$ changes, only RCP waves can acquire geometric phase changes ($\beta_2$-geometric phase, for short) while RCP waves does not. Such spin-decoupled geometric phase can be readily combined with other geometric phases, such as conventional spin-coupled P-B phase by rotating meta-atoms. When the whole meta-atom is rotated by an angle α, LCP and RCP waves will gain opposite-signed geometric phases (α-geometric phase, for short) additionally.

To further verify this design philosophy, three phase-gradient metasurse prototypes were designed based on α-geometric phase, $β_1$-geometric phase and (α+$β_1$)-geometric phase. The anomalous reflection angle $θ_r$ for reflective phase-gradient metasurface can be calculated by $\sin θ_r = -\frac{\lambda}{2\pi} \times \frac{\Delta\varphi}{P}$, where P is the period of the meta-atom, Δφ is the phase step between each two adjacent meta-atoms. As shown in Fig 5(a), by changing $β_1$, six meta-atoms with a reflection phase step of 60° can be obtained under normal incidence, so as to form a phase gradient only for LCP waves. The lower two panels in Fig 5(a) show the simulated normalized far-field directions under RCP (left panel) and LCP (right panel) waves. It is clearly shown that under LCP incidence, the main lobe of reflected waves is deflected and the deflected angle is consistent with theoretical calculation. In contrast, under RCP incidence, the main lobe of reflected waves is not deflected, still located at the normal direction. For the sake of comparison, we also designed a phase gradient metasurface based on α-geometric phase by simply rotating the meta-atoms in sequence, as shown in Fig 5(b). $β_1$ is kept unchanged and the six identical meta-atoms are rotated by α=30° one by one in sequence. In this way, LCP and RCP waves will obtain ±2α P-B phase steps, respectively. That is, the phase gradients are opposite in directions for LCP and RCP waves and such a geometric phase like this is naturally spin-coupled. Unlike the far-field distribution in Fig.5(a), the gradient metasurface in Fig.5(b) deflects LCP and RCP waves towards opposite directions with equal deflected angles, which is consistent with theoretical expectations.

Since the generation mechanism of the spin-decoupled $β_1$-geometric phase and the spin-coupled α-geometric phase are different, it comes naturally that we can combine them so as to enhance the DOF in phase modulation of EM waves. Under such a consideration, we also designed a metasurface based on the combination of $β_1$-geometric phase and α-geometric phase, which can realize different functions for LCP and RCP waves. According to theoretical analysis, the geometric phase obtained by LCP and RCP waves are, respectively.

$$\begin{cases} \phi_L = 2(β_1 - 40) + 2α, & 40° \leq β_1 \leq 220° \\ \phi_R = -2α, & 0° \leq α \leq 180° \end{cases} \quad (3)$$

where $\phi_L$ and $\phi_R$ are the phase step imparted to LCP and RCP waves by the metasurface, respectively. Specifically, we designed a metasurface based on (α+$β_1$)-geometric phases simultaneously with $\Delta\varphi_L$=-45° and $\Delta\varphi_R$=-60°. According to the phase calculation formula, parameters of the required units can be obtained, with a total of 24 meta-atoms, as shown in Fig 5(c). The scattering pattern at the central frequency 11GHz and the normalized far-field scattering spectrum are obtained for LCP and RCP waves. It can be seen that compared with the result in Fig.5(a) and (b), the waves under LCP and RCP are deflected towards the same side, but with different deflected angles. At 11.0GHz, LCP waves are deflected by 20.0°, while RCP waves are deflected by 27.0°, which is consistent with the theoretical calculation. It can be found from the spectrum diagram that deflected reflection occurs in a wide band from 8.0GHz to14.0GHz, with high efficiency.

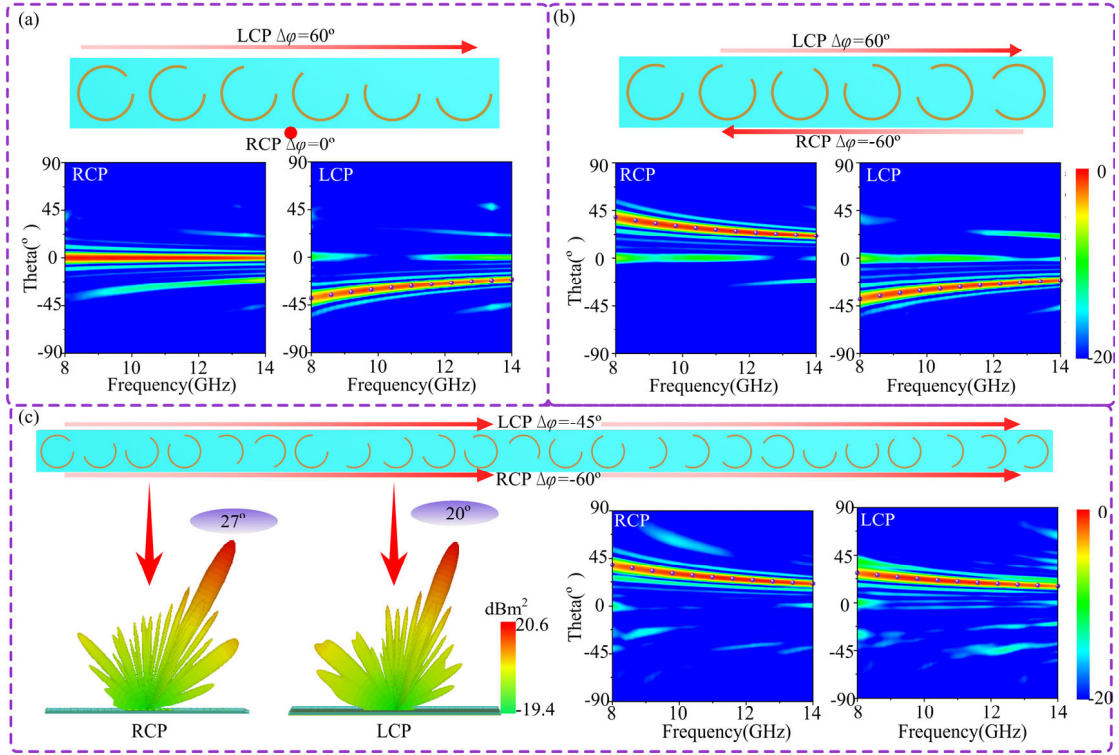

Fig.5 Design and simulated results of the three prototypes of phase-gradient metasurface based on geometric phases: (a) the completely spin-decoupled metasurface based on $\beta_1$-geometric phase. (b) the spin-coupled metasurface based on $\alpha$-geometric phase. (c) the spin-decoupled metasurface based on $(\alpha+\beta_1)$-geometric phase.

Furthermore, linearly polarized (LP) waves can be regarded as the superposition of LCP and RCP waves. For the metasurface based on $(\alpha+\beta_1)$-geometric phase, LCP and RCP waves can be deflected towards the same side with different deflected angles. When LP waves are incident upon the metasurface, two beams of reflected waves will be generated. Fig 6(a) and (b) show normalized reflection spectra and three-dimensional scattering patterns of the two metasurfaces based on $\alpha$-geometric phase and spin-decoupled $(\alpha+\beta_1)$-geometric phase under normal incidence of $x$-polarized wave, respectively. It can be found that the simulated reflection angles are consistent with the simulation results. To further verify the spin-independent phase control performance, the protype based on $(\alpha+\beta_1)$-geometric phase was fabricated and measured. Fig 6(c) shows the measurement setup. Two $x$-polarized antennas are used as the receiving and transmitting antennas for the measurement. Panels (ii)-(iv) in Fig 6 (c) plot measured results at different frequencies (8.0GHz, 11.5GHz and 13.0GHz). It can be found that the measured results are consistent with simulated ones. This verifies the validity of our method of obtaining spin-decoupled geometric phase by engineering the surface current patch on meta-atoms.

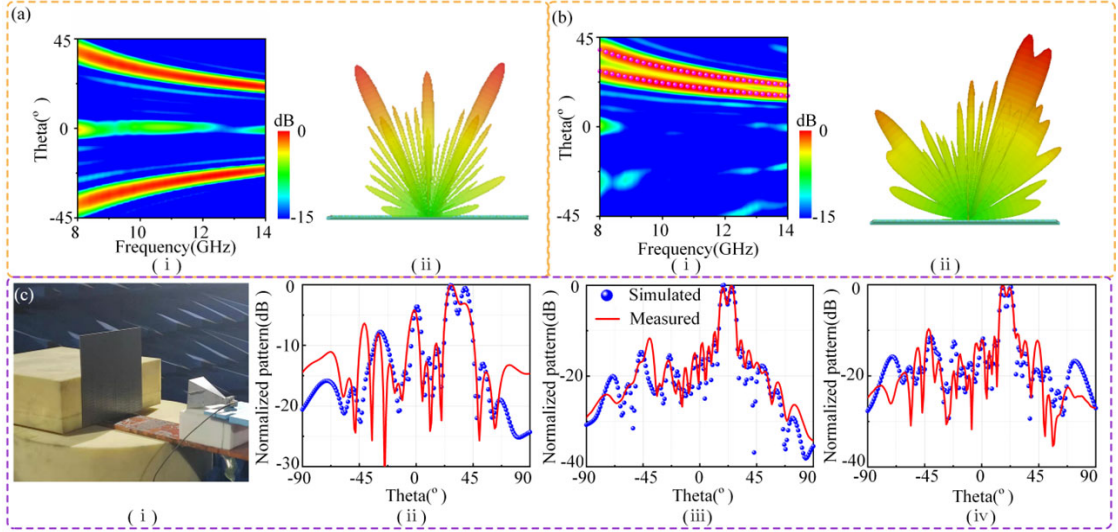

Fig 6 Beam splitting and deflection under LP waves: (a) reflection performance (i) and far-field scattering pattern (ii) of the metasurface based on $\alpha$-geometric phase at 11.5GHz. (b) reflection performance (i) and far-field scattering pattern (ii) of the metasurface based on $(\alpha+\beta_1)$-geometric phase at 11.5GHz. (c) The metasurface prototype based on $(\alpha+\beta_1)$-geometric phase in the measurement system (i) and the measured far-field pattern results at (ii) 8.0GHz, (iii) 11.5GHz and (iv) 13.0GHz.

## 4. Conclusions

In conclusions, we explore completely spin-decoupled geometric phases of metasurfaces by engineering the surface current patch of meta-atoms. The generation mechanism of spin-coupled geometric phase obtained by rotating meta-atoms is firstly analyzed and it is shown that the essence lies in the rotation of induced surface current. It is straightforward to rotate the surface current patch by rotating the meta-atom as a whole. However, geometric phases of such are always spin-coupled, since both the surface currents induced by LCP and RCP waves are rotated by the same angle. Therefore, by rotating the meta-atom, geometric phases with the same magnitude but opposite signs can be imparted to LCP and RCP waves. In fact, rotation of surface current doesn't not necessarily have to resort to rotating the meta-atom. By engineering the start points of surface current paths under LCP and RCP waves, rotation of surface current can also be achieved conveniently. More importantly, the start point of surface current under LCP waves can be made different from that under RCP waves. In this way, the geometric phases imparted to LCP and RCP waves by the metasurface can be completely decoupled. Prototypes were designed, fabricated and measured. Both the simulation and measurement results verify this design philosophy. Our work provides a robust method of obtaining spin-independent geometric phases and further adds up to the DOF of metasurface in manipulating EM waves.

## Acknowledgment

This work was supported by the National Natural Science Foundation of China (Nos. 62101588, 61971435); the Young Talent Fund of Association for Science and Technology in Shaanxi, China(No. 20220102).